\documentclass[letterpaper,nobalancelastpage]{revtex4}

\usepackage{graphicx}
\usepackage{amsmath}
\usepackage{amssymb}
\usepackage{units}
\usepackage{array}

\begin{document}

\title{Long range Coulomb interaction in the ground state of bilayer
graphene}

\author{D. S. L. Abergel}
\email{abergel@cc.umanitoba.ca}
\author{Tapash Chakraborty}
\affiliation{Department of Physics and Astronomy, University of
Manitoba, Winnipeg MB, R3T 2N2, Canada.}

\begin{abstract}
We report on our studies of interacting electrons in bilayer graphene in
a magnetic field.
We demonstrate that the long range Coulomb interactions between
electrons in this material are highly important. 
We show that in the unbiased bilayer (where both layers are at the same
electrostatic potential), the interactions can cause mixing of Landau
levels in moderate magnetic fields. 
For the biased bilayer (when the two layers are at different
potentials), we demonstrate that the interactions are responsible for a
change in the total spin of the ground state for half-filled Landau
levels in the valence band.
\end{abstract}

\maketitle

%%%%%% Outline
% Monolayer
Monolayer graphene is a two-dimensional hexagonal crystal of carbon
atoms which exhibits a number of physical and electronic properties that
have made it the subject of intense study since it was first isolated in
2004 \cite{novoselov:sci306}. 
The existence of a gapless, conical low energy quasiparticle spectrum
$E=\hbar v_F k$ near the $K$ points of the Brillouin zone (with the
slope of the dispersion relation defined by the Fermi velocity $v_F
\approx c/300$) has lead to the possibility of realizing relativistic
effects in a table-top solid state experiment \cite{geim:natmat6}.
Additionally, the unusual chiral nature of the charge carriers was
confirmed by the observation of an anomalous integer quantum Hall effect
\cite{novoselov:nature438} with Landau level (LL) spectrum $E_n =
\sqrt{n} \hbar v_F/\lambda_B$, including a LL at zero energy and
$\sqrt{B}$ dependence of the spectrum on the magnetic field ($\lambda_B$
is the magnetic length).
From a more practical point of view, the unusually high electron
mobility and conductance of graphene at room temperature
\cite{tan:prl99} makes this a very promising material for the
fabrication of electronic devices.

\begin{figure}
	\centering
	\includegraphics{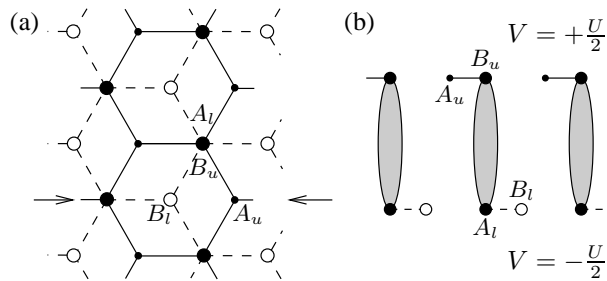}
	\caption{The lattice structure of bilayer graphene. The upper
	(lower) lattice is shown by solid (dashed) lines. The $A_u$ ($B_l$)
	sublattices are shown with small, filled (large, open) dots; the
	$B_u$-$A_l$ dimers by large, filled dots. (a) The
	top-down view; (b) the side-on view projected between the two
	arrows in (a). 
	\label{fig:lattice}}
\end{figure}

% Bilayer - what it is
Bilayer graphene (BLG) \cite{mccann:ssc143} consists of a pair of
monolayers bonded by relatively weak dimer bonds perpendicular to the
plane of the monolayer sheets (see Fig. \ref{fig:lattice}). 
In this material, both the conduction and valence bands have low
energy structure consisting of two quadratic branches seperated by the
energy associated with the dimer bond, $\gamma_1$.
The lower conduction band and upper valence band are degenerate at the
$K$ points. 
The existence of chiral charge carriers with a Berry's phase of
$2\pi$ was confirmed in the observation of the integer quantum Hall
effect \cite{novoselov:natphys2} where the low energy LL spectrum is
approximately linear in the field with $\epsilon_n \simeq \pm \hbar\omega_c
\sqrt{n(n-1)}$ for $n\geq0$ where $\omega_c$ is the cyclotron frequency. 
This spectrum includes a doubly degenerate LL at zero energy
\cite{mccann:prl96}.

% Description of the effect of the gap - single particle
It has been predicted theoretically \cite{mccann:prl96, mccann:prb74,
pereira:prb76} and observed experimentally (first by doping of the
bilayer \cite{ohta:science313}, and then by applying electrostatic gates
\cite{castro:prl99}) that a gap can be induced in the low energy band
structure of BLG by breaking the symmetry between the two
layers. Switching off of the conduction current (and hence, proof of
principal for switching devices made from so-called `biased bilayer
graphene') by sweeping the Fermi energy through the gapped region has
been observed at low temperatures \cite{oostinga:natmat7}, and this
development has lead to a surge in interest in this system.

% What we do - interactions in monolayer and ungapped bilayer
The single particle theory of this material is well known
\cite{mccann:ssc143, mccann:prl96, mccann:prb74, pereira:prb76}, it has
been shown that the electron-electron interactions are significant in
monolayer graphene \cite{iyengar:prb75, apalkov:prl97}.
The Coulomb interaction has been studied in the unbiased (and hence,
ungapped) bilayer \cite{nilsson:prb73}, while the biasing potential was
considered in the context of a ferromagnetic transition due to
short-range interactions in the mean-field approximation at zero
magnetic field \cite{castro:prl100}. 
Also, certain collective modes leading to an intra-Landau level
cyclotron resonance have been predicted in the presence of a magnetic
field within Hartree-Fock theory \cite{barlas:prl101}. 
However, the effect of the long range Coulomb interaction on the ground
state of the biased system in a strong magnetic field has not been
systematically investigated, and we address this problem in the current
Letter.

% Our results
We find that the long range nature of the Coulomb interaction makes
significant changes to the properties of the low energy charge carriers
for BLG in a magnetic field. 
Specifically, the interactions are significantly stronger for 
electrons in the lowest LL, and this manifests itself in an
observable way. 
It allows the possiblity of mixing of the LLs which were well-seperated
in energy when the interactions were not considered.  By calculating the
explicit form of the ground state wave function, we show how this mixing
fundamentally changes the nature of the ground state in the biased
bilayer, and how this change manifests itself in the total spin of the
ground state of the interacting system for half-filled LLs.

% Technical details
% Intro to bilayer
In BLG, each sheet contains two inequivalent triangular
sublattices of carbon atoms which we label $A_u$, $B_u$, $A_l$, and
$B_l$ for the $A$ and $B$ sublattices of the upper and lower layers
respectively (see Fig. \ref{fig:lattice}). 
In the Bernal stacking arrangement, the inter-layer bonds consist of
dimers formed from atomic orbitals associated with the $B_u$ and $A_l$
sublattices. The energy associated with this bond is denoted $\gamma_1$
and throughout this Letter we assume that $\gamma_1 = 0.4\unit{eV}$. 
We allow for a static electric potential $U$ to be applied between the
upper and lower layers, so that the upper layer has potential $U/2$, and
the lower layer $-U/2$. In a strong magnetic field we can write the
tight-binding Hamiltonian using the four component basis
$\{A_u,B_l,A_l,B_u\}$ with $\xi=+1$ in the $K$ valley, and the basis
$\{B_l,A_u,B_u,A_l\}$ with $\xi=-1$ in the $K'$ valley as
\cite{mccann:prl96}
\begin{equation}
	\mathcal{H}_0 = \begin{pmatrix}
		\frac{\xi U}{2} & 0 & 0 & \xi v_F \pi^\dagger \\
		0 & -\frac{\xi U}{2} & \xi v_F \pi & 0 \\
		0 & \xi v_F \pi^\dagger & -\frac{\xi U}{2} & \gamma_1 \\
		\xi v_F \pi & 0 & \gamma_1 & \frac{\xi U}{2} \end{pmatrix}
	\label{eq:Ham}
\end{equation}
where $\pi$ and $\pi^\dagger$ are the operators corresponding to
electron hops between neighbouring atoms (in opposite sublattices in the
same layer).
The eigenvalues $\varepsilon_n^\xi$ of Hamiltonian \eqref{eq:Ham} are found
from the quartic polynomial equation \cite{pereira:prb76}
\begin{equation}
	\left[ 2n - \left(\xi\delta + \varepsilon_n^\xi \right)^2 \right]
		\left[ 2(n+1) - \left(\xi\delta 
		- \varepsilon_n^\xi \right)^2 \right] \\
	= t^2 \left[ \left(\varepsilon_n^{\xi}\right)^2 - \delta^2 \right]
	\label{eq:eval}
\end{equation}
where $n\in \mathbb{Z}^+$. The energies $t\propto \gamma_1$, $\delta
\propto \frac{U}{2}$ and $\varepsilon_n^\xi$ are measured in units of
$\frac{\hbar v_F}{\lambda_B}$. Additionally, we denote the band of a
particular LL by placing a `$+$' (`$-$') after the level index
for the conduction (valence) band. The wave functions associated with
the LLs of this spectrum for $n\geq1$ are given by
\begin{equation}
	\psi_{n\pm}^\xi = e^{iky} \left( a_{n\pm}^\xi \varphi_{n+1}, \,
	b_{n\pm}^\xi \varphi_{n-1}, \,
	c_{n\pm}^\xi \varphi_n, \,
	d_{n\pm}^\xi \varphi_n \right)^T
	\label{eq:evec}
\end{equation}
where the functions $\varphi_n$ for $n\geq 0$ are the magnetic
oscillator functions in the Landau gauge and the coefficients are
defined so that the overall wave function is normalized to unity. 
There are also levels with $n=0\pm$, which have wave functions 
$\psi_{0+}^K = e^{iky} \left( \varphi_0, 0, 0, 0 \right)$, and
$\psi_{0-}^{K'} = e^{iky} \left( \varphi_0, 0, 0, 0 \right)$ with
$\varepsilon=\pm\delta$; and $\psi_{0+}^{K'}$ and $\psi_{0-}^K$ are
defined by Eqns. \eqref{eq:eval} and \eqref{eq:evec} with the
appropriate substitutions for $n$ and $\xi$. When $U=0$, these four
states are degenerate yielding the eight-fold degeneracy (including the
factor of 2 for spin) seen in the integer quantum Hall effect in BLG
\cite{novoselov:natphys2}.  Note that for a sufficiently strong
inter-layer potential, the single
particle energies described here can cross at certain magnetic field
strengths.
We include the fermionic properties of the electrons by constructing
Slater determinants for the non-interacting many body state wave
functions.

% Coulomb interaction and exact diagonalization
To include the effects of the long-range Coulomb interaction we consider
the Hamiltonian
\begin{equation}
	\mathcal{H}_{\mathrm{Coul}} = \frac{1}{2} \sum_{i\neq j}
	\frac{e^2}{\epsilon |\vec{r}_i-\vec{r}_j|}
\end{equation}
where the vectors $\vec{r}_{i,j}$ label the positions of the electrons,
and $\epsilon$ is the dielectric constant of graphene. For graphene
mounted on an SiO$_2$ substrate, $\epsilon_{\text{SiO}_2} = 2.5$
\cite{graphene-dielectric}.
Other methods of isolating graphene could be considered by changing the
value of this constant.
The effect of the interaction term is to mix the non-interacting states
into some linear combination which minimizes the total energy of the
system.  Our analysis is conducted by employing the exact
diagonalization scheme \cite{exactdiag} in which we calculate the
linear combination of non-interacting basis states which gives the
ground state of the Hamiltonian $\mathcal{H} = \mathcal{H}_0 +
\mathcal{H}_\mathrm{Coul}$. 
This method entails dividing the infinite sheet into rectangular cells
of dimension $L_x\times L_y$ \cite{asprat} and applying periodic
boundary conditions to the wavefunctions at the edges of each cell.
The matrix elements of the interaction over the single particle states
described above are evaluated exactly by calculating the integrals over
the spatial extent of the cell numerically.
The interaction between the cells is taken into account by adding the
Madelung energy of a charged lattice \cite{madelung}. 

The number of single particle states included in the Hilbert
space from which the non-interacting many body basis states are
constructed is determined as follows. 
There are four relevant quantum numbers: The LL index $n$, the
valley $\xi$, the spin and the momentum $k=\pi m/L_x$. 
The values of the momentum are fixed when the boundary conditions are
applied to the cell, and are labelled by $0\leq m \leq M-1$ with $M=L_x
L_y/(2\pi\lambda_B^2)$.
Which LLs we select is governed by the details of the system
we wish to model, and $M$ is set by computational restraints.

In order to reduce the size of the many body system (and so to improve
the speed of calculation), we observe that the Hamiltonian conserves the
total momentum $\mu=\sum_{i=1}^N m_i \mod M$. Therefore, we can perform
the diagonalization seperately for each value of $\mu$ which reduces the
basis size to approximately the 1/$M$th part. Finally, since there is no
spin-dependent term in the Hamiltonian, the total projection of spin on
the $z$ axis $S_z$ is also a good quantum number. We can fix $S_z$ to
its minimum value (which is $0$ for even $N$ and $\frac{1}{2}$ for odd
$N$) whilst still being able to recover all eigenstates of the total
spin operator $S^2$ \cite{apalkov:prl97}.

We consider two different cases:
Firstly we demonstrate the strength of the interaction by calculating
the shift in the energy of each LL due to interactions for $U=0$
(an unbiased bilayer).
Then we examine the system where the filling factor is negative, the
inter-layer potential is sizeable and the magnetic field strong. In this
case, we observe changes in the total spin of the ground state as a
function of $U$ and the magnetic field strength $B$.

\begin{table}[tb]
	\centering
	\setlength{\extrarowheight}{2pt}
	\begin{tabular}{|c|| >$c<$ | >$c<$ | >$c<$ | >$c<$ |}
	\hline
	$\nu$        &    -3   &    -2   &    -1   &    0    \\
	Energy shift & -0.6443 & -0.6443 & -0.6443 & -0.6443 \\ \hline
	$\nu$        &     1   &     2   &     3   &    4    \\
	Energy shift & -0.6316 & -0.6222 & -0.6148 & -0.6085 \\ \hline
	\end{tabular}
	\caption{Energy shift per electron due to the Coulomb interaction
	for integer filling factors in the $0\pm$ LL for $U=0$. 
	Energy units are $e^2/(\epsilon\lambda_B)$,
	the number of momentum states $M=3$, 
	and the magnetic field $B=3\unit{T}$.  
	\label{tab:zb-n0}}
\end{table}

We model the unbiased bilayer near half-filling by takng a single
particle Hilbert space consisting of electrons in the $0+$ and $0-$ LLs
with all possible spin and valley states at $U=0$. Each integer value of
the filling factor $\nu$ is
simulated by taking the number of electrons $N=(\nu+4)M$, and we have
$M=3$.  Table \ref{tab:zb-n0} shows the results of diagonalizing the
resulting many body Hamiltonian and evaluating the change in energy from
the non-interacting ground state for integer filling factors.
We notice that the energy shift per electron reduces slightly as the LL
fills.

In Table \ref{tab:zb-nfin}, we show the energy shift due to
the Coulomb interaction for electrons in higher LLs
(\textit{i.e.} for levels with $n\geq 1$) in the conduction band. 
We have taken a single particle Hilbert space consisting of all spin and
valley states within one LL.
The filling factor $\nu_n$ within LL $n$ can range between $0$
(corresponding to an empty level) and $4$ (a filled level), so that
$\nu_n=4$ and $\nu_{n+1}=0$ describe the same overall filling factor. 
The number of electrons is set by $N=\nu_n M$, and in order to allow
direct comparison with the lowest LL we restrict ourselves to $M=3$.
Table \ref{tab:zb-nfin} shows that the energy associated with the
interaction of electrons is very similar in each of the higher LLs, and
that the interaction energy per particle is slightly reduced as the
LL is filled.
We have verified that the results are identical in the valence band.

\begin{table}[tb]
	\setlength{\extrarowheight}{2pt}
	\centering
	\begin{tabular}{ |c| |>$c<$ |>$c<$ |>$c<$ |>$c<$ |}
	\hline
	Landau level $n$ &    1+   &    2+   &    3+   &    4+   \\ \hline
	$\nu_n = 1$      & -0.4766 & -0.5001 & -0.5242 & -0.5160 \\
	$\nu_n = 2$      & -0.4705 & -0.4880 & -0.5176 & -0.5110 \\
	$\nu_n = 3$      & -0.4645 & -0.4759 & -0.5111 & -0.5061 \\
	$\nu_n = 4$      & -0.4584 & -0.4638 & -0.5046 & -0.5012 \\ \hline
	\end{tabular}
	\caption{Energy shift per electron due to the Coulomb interaction in
	LL $n\geq 1$ at filling factor $\nu_n$. Energy units are
	$e^2/(\epsilon\lambda_B)$, the
	number of momentum states $M=3$, and the magnetic field
	$B=3\unit{T}$. 
	\label{tab:zb-nfin}}
\end{table}

\begin{figure}[bt]
	\centering
	\includegraphics[]{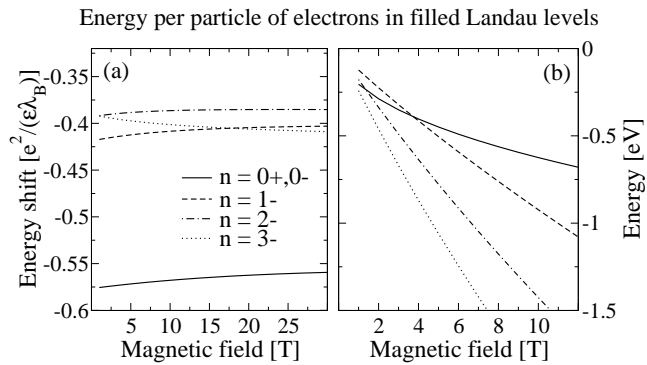}
	\caption{(a) The energy shift per electron of filled LLs.
	(b) The absolute energy per electron of filled LLs
	showing the crossing between the $n=0\pm$ degenerate level
	and the higher LLs in the valence band.
	In both plots $U=0$, and $M=5$.
	\label{fig:flplots}}
\end{figure}

% Interpretation of U=0 results
% Table 1 - lowest Landau level
% Table 2 - higher Landau levels
% Wider implications - possiblity of level mixing
Together, Tables \ref{tab:zb-n0} and \ref{tab:zb-nfin} show that the
effect of the long range electron-electron interactions is considerable,
and that for this value of the magnetic
field ($B=3\unit{T}$) the shift in the higher LLs is only about
two-thirds that of the lowest LL. 
In Fig. \ref{fig:flplots} we show the energy shift and absolute energy
of filled LLs as a function of the magnetic field. The
strength of the interaction scales with $e^2/(\epsilon\lambda_B) \propto
\sqrt{B}$ with a roughly constant coefficient, while the LL spacing goes
with $\omega_c \propto B$, so for lower values of the field, the
$n=0\pm$ level crosses the $2-$ and $1-$ levels as shown in Fig.
\ref{fig:flplots}(b).
The data shown here were calculated with $\epsilon=2.5$, modelling
graphene \cite{graphene-dielectric} deposited on an SiO$_2$ substrate.
For suspended graphene (where $\epsilon\approx 1$), it is conceivable
that the effect of the interaction would be even stronger.
Additionally, the effect of the inter-layer potential is to bring
together the valence band LLs with low index, so for finite bias it is
plausible that the interactions will cause significant mixing between
these levels.

\begin{figure}[tb]
	\centering
	\includegraphics[]{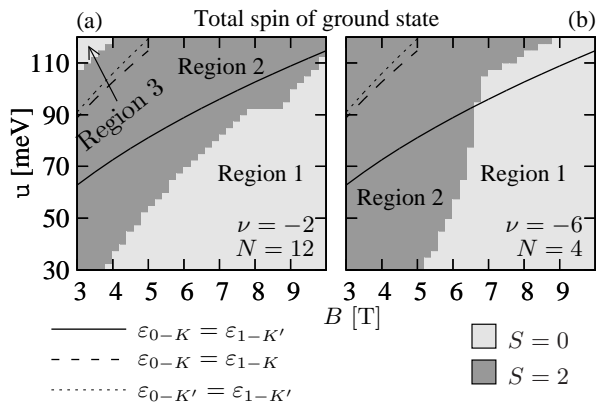}
	\caption{The total spin of the ground state of the (a) $\nu=-2$ and
	(b) $\nu=-6$ systems. $M=2$ and $N=(\nu+8)M$. 
	The lines show the crossing points of the single particle states.
	The graining is due to the finite interval between data	points.
	\label{fig:spins}}
\end{figure}

% Introduction to spin changes
Now we turn our attention to the system with an inter-layer potential
applied, so that $U\neq 0$. 
With a finite gap between the $0+$ and $0-$ levels and non-zero
filling factor, it is possible to consider the negatively-doped system
by taking only those single particle states which are in the valence
band. Therefore we select the single particle states from which to
form the Slater determinants describing the non-interacting basis states
by taking all spin and valley states of the $0-$ and $1-$ LLs.
We have $M=2$ and the number of electrons is related to the
filling factor by $N=(\nu+8)M$.  Diagonalizing this system for
half-filled LLs (so for $\nu=-6$ and $\nu=-2$), and
calculating the expectation value of the total spin operator $S^2$ over
the resulting ground state as a function of the magnetic field and the
inter-layer bias gives the data shown in the plots in Fig.
\ref{fig:spins} \cite{valleyspin}.
We have also superimposed lines which represent the
energy at which the single particle energy levels cross, as labelled in
the legend.
% Discussion of results
The plots show that there is an abrupt change in the total spin of the
ground state, and a range of parameters where there is a non-zero
polarization of the spin. 
This transition is not directly related to the crossings of the single
particle states, since the position and slope of the transitions do not
match the corresponding lines which we have superimposed on the plots.
This effect is therefore purely due to the Coulomb interaction, and in
particular to the exchange contribution which acts to minimize the
energy of spin-polarized many body states.

\begin{figure}[bt]
	\centering
	\includegraphics[]{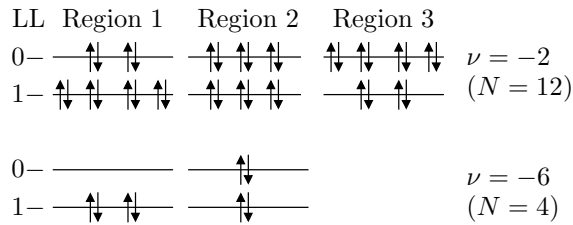}
	\caption{The occupancy of the single electron states in the
	intaracting many body ground state for each region of the plots in
	Fig. \ref{fig:spins}. The total $z$-projection of the spin is
	always zero since we fix this when forming the non-interacting many
	electron basis states.
	\label{fig:gsocc}}
\end{figure}

Figure \ref{fig:gsocc} shows the occupation of the single particle
levels in the interacting many body ground state. 
For simplicity, we display only the LL index of the states.
The actual ground state is a coherent combination of several 
non-interacting basis states, where the combination of LL
indices is consistent but different arrangements of momentum and valley 
states each come with identical prefactors in the linear combination.
This figure shows that in the lower-right region of
the parameter space considered, the electrons occupy as
many of the $1-$ states as possible. As the parameters change to the
upper-left region, the $0-$ levels become successively more populated.
This demonstrates the fundamental effect of the inter-layer potential
and the electron-electron interactions on the nature of the ground
state.

% Summary
In conclusion, we have shown that the long range Coulomb interaction
between electrons plays an important r\^ole near the Dirac point in
BLG. In the unbiased case, the interactions will
cause electrons
in the doubly-degenerate lowest LL to be lower in energy than those in
the $1-$ level for small-to-moderate magnetic fields. If an inter-layer
bias is applied to split the valence and conduction bands, the
electron-electron interactions precipitate a transition in the total
spin of the ground state of half-filled LLs for certain ranges of
parameters. This effect will have fundamental implications for the
design of devices made from this material.

We acknowledge financial support from the Canada Research Chairs
Program, and the NSERC Discovery Grant.

% Bibliography


\begin{thebibliography}{}
\bibitem{valleyspin} We have also calculated the expectation of the
total valley quantum number for each of these systems and find that it
is constant with a value equal to 2.
\end{thebibliography}
\end{document}